\documentclass[sigconf,screen]{acmart} 

\acmConference[ICSE 2024]{46th International Conference on Software Engineering}{April 2024}{Lisbon, Portugal}

\AtBeginDocument{%
  \providecommand\BibTeX{{%
    \normalfont B\kern-0.5em{\scshape i\kern-0.25em b}\kern-0.8em\TeX}}}

\newcommand{\midsepremove}{\aboverulesep = 0mm \belowrulesep = 0mm}
    \midsepremove
    \newcommand{\midsepdefault}{\aboverulesep = 0mm \belowrulesep = 0mm}
    \midsepdefault

\usepackage{graphicx}
\usepackage{xcolor}
\graphicspath{{figs/}}
\usepackage{makecell}
\usepackage{tcolorbox}
\tcbuselibrary{skins, breakable, theorems}
\usepackage{diagbox}
\usepackage{hyperref}
\usepackage{caption}
\usepackage{multirow}
\usepackage{subfigure}
\usepackage{threeparttable}
\makeatletter  
\newif\if@restonecol  
\makeatother  
\usepackage[linesnumbered,ruled,vlined]{algorithm2e}
\usepackage{algpseudocode}  
\usepackage{enumitem}
\usepackage{calc}
\usepackage{booktabs}
\usepackage{colortbl}

\definecolor{lightgray}{HTML}{eeeeee}
\definecolor{tab_red}{rgb}{1,0.76,0.71}

\newcommand{\tool}{\textsc{MultiTest}}

\newcommand{\rqone}{RQ1. How effective is the {\tool} at synthesizing realistic multi-modal data?}
\newcommand{\rqtwo}{{RQ2. How effective is the {\tool} at generating error-reveling tests?}}
\newcommand{\rqthree}{{RQ3. How effective is the {\tool} at guiding the improvement of a SUT through retraining?}}

\usepackage{color}

\definecolor{grayblue}{rgb}{0.2,0.29,0.79}
\definecolor{darkgreen}{rgb}{0.2,0.55,0.1}
\definecolor{violet}{rgb}{0.54,0.17,0.88}

\newcommand{\myadd}[1]
{
   {\noindent\color{red}\bf #1}
}
\newcommand{\mydel}[1]
{
   {\noindent\color{blue} \sout{#1}}
}

\newcommand{\ma}[1]
{
   {\noindent\color{red}\bf [#1]$_{\scriptscriptstyle\textit{ma}}$}
}

\newcommand{\feng}[1]
{
   {\noindent\color{red}\bf [#1]$_{\scriptscriptstyle\textit{feng}}$}
}
\newcommand{\gao}[1]
{
   {\noindent\color{grayblue}\bf [#1]$_{\scriptscriptstyle\textit{gao}}$}
}

\newcommand{\wang}[1]
{
   {\noindent\color{violet}\bf [#1]$_{\scriptscriptstyle\textit{wang}}$}
}

\newcommand{\includeAuthorComments}[1]
{
   \ifthenelse{\equal{#1}{0}}
   {
      \renewcommand{\feng}[1]
      {
         {} 
      }
      \renewcommand{\gao}[1]
      {
         {} 
      }
      \renewcommand{\wang}[1]
      {
         {} 
      }
       \renewcommand{\ma}[1]
      {
         {} 
      }
        \renewcommand{\myadd}[1]
      {
         {} 
      }
        \renewcommand{\mydel}[1]
      {
         {} 
      }
      
   }{}
}

\definecolor{royalblue}{RGB}{65,105,225}

\newcommand{\reviseref}[0]{\color{black}}
\newcommand{\reviseline}[1]{\textcolor{black}{#1}}

\definecolor{tab_revise}{rgb}{1,0,0}

\copyrightyear{2024}
\acmYear{2024}
\setcopyright{acmlicensed}\acmConference[ICSE '24]{2024 IEEE/ACM 46th International Conference on Software Engineering}{April 14--20, 2024}{Lisbon, Portugal}
\acmBooktitle{2024 IEEE/ACM 46th International Conference on Software Engineering (ICSE '24), April 14--20, 2024, Lisbon, Portugal} 
\acmDOI{10.1145/3597503.3639191} 
\acmISBN{979-8-4007-0217-4/24/04}

\begin{document}

\title{{\tool}: Physical-Aware Object Insertion for Testing Multi-sensor Fusion Perception Systems}

\author{Xinyu Gao}
\authornote{Both authors contributed equally to this work.}
\affiliation{
  \institution{State Key Laboratory for Novel Software Technology}
\country{Nanjing University, China}
}

\author{Zhijie Wang}
\authornotemark[1]
\affiliation{
  \institution{University of Alberta}
  \city{Edmonton}
\country{Canada}
}

\author{Yang Feng}
\authornote{Yang Feng, Lei Ma, and Zhenyu Chen are the corresponding authors.}
\affiliation{
  \institution{State Key Laboratory for Novel Software Technology}
\country{Nanjing University, China}
}

\author{Lei Ma}
\authornotemark[2]
\affiliation{%
  \institution{The University of Tokyo, Japan}
  \city{}
  \state{}
  \country{University of Alberta, Canada}
  }

\author{Zhenyu Chen}
\authornotemark[2]
\affiliation{
  \institution{State Key Laboratory for Novel Software Technology}
\country{Nanjing University, China}
}

\author{Baowen Xu}
\affiliation{
  \institution{State Key Laboratory for Novel Software Technology}
\country{Nanjing University, China}
}

\begin{abstract}
\textbf{Multi-sensor fusion} stands as a pivotal technique in addressing numerous safety-critical tasks and applications, e.g., self-driving cars and automated robotic arms. With the continuous advancement in data-driven artificial intelligence (AI), MSF's potential for sensing and understanding intricate external environments has been further amplified, bringing a profound impact on intelligent systems and specifically on their perception systems.
Similar to traditional software, adequate testing is also required for AI-enabled MSF systems. 
Yet, existing testing methods primarily concentrate on single-sensor perception systems (e.g., image-based and point cloud-based object detection systems). There remains a lack of emphasis on generating multi-modal test cases for MSF systems. 

To address these limitations, we design and implement {\tool}, a fitness-guided metamorphic testing method for complex MSF perception systems. 
{\tool} employs a physical-aware approach to synthesize realistic multi-modal object instances and insert them into critical positions of background images and point clouds.
A fitness metric is designed to guide and boost the test generation process.
We conduct extensive experiments with five SOTA perception systems to evaluate {\tool} from the perspectives of: (1) generated test cases'  realism, (2) fault detection capabilities, and (3) performance improvement.
The results show that {\tool} can generate realistic and modality-consistent test data and effectively detect hundreds of diverse faults of an MSF system under test. Moreover, retraining an MSF system on the test cases generated by {\tool} can improve the system's robustness. Our replication package and synthesized testing dataset are publicly available at \href{https://sites.google.com/view/msftest}{https://sites.google.com/view/msftest}.

\end{abstract}




\maketitle

\section{Introduction}
%
Multi-sensor fusion (MSF) plays a vital role in various intelligent machines and software systems. The recent rapid advancements in data-driven artificial intelligence (AI) and sensor technologies further propelled progress in the development of MSF-based systems. Nowadays, prominent industrial-level systems, such as OpenPilot~\cite{openpilot}, commonly rely on multi-sensor fusion strategies to overcome the inherent limitations of individual sensors and enhance overall system performance~\cite{survey_feng2020deep}. 
Consequently, MSF-based perception systems have found widespread applications in diverse industrial domains and safety-critical use cases, including self-driving cars~\cite{survey_cui2021deep}, unmanned aerial vehicles~\cite{gupta2022simultaneous}, and robotic systems~\cite{debeunne2020review}. 


Despite the rapid progress, AI-enabled perception systems, similar to any traditional software, can still yield incorrect prediction results, which can further lead to incorrect system behavior. Oftentimes the incorrect behavior could result in severe accidents and losses in safety-critical contexts, e.g., autonomous driving. For instance, a Tesla self-driving car failed to distinguish a white truck against a bright sky~\cite{TeslaAccident2}, causing a fatal collision.

To improve the overall quality of AI-enabled perception systems, software engineering and machine learning researchers have proposed a few quality assurance techniques for the different development stages, e.g., testing~\cite{wang2020metamorphic,10172508, guo2022lirtest,xie2022towards} and debugging~\cite{kim2020programmatic,leclerc20223db}. Among these approaches, testing has emerged as a proven and efficient method to assess the potential risks of deploying an AI-enabled perception system in real-world scenarios. A typical testing workflow takes a small set of tests (e.g., images) as seeds and generates more challenging test cases based on seeds. Existing testing techniques for AI-enabled perception systems usually leverage natural/adversarial perturbations to synthesize new test data~\cite{guo2022lirtest,xie2022towards}. However, they mostly focus on testing single-sensor (e.g., camera or LiDAR) perception systems. Yet, little has been done from the perspective of testing MSF-based systems~\cite{gao2023benchmarking}. A systematic survey from practitioners in autonomous driving shows that there comes an urgent need for multi-modal data synthesis techniques~\cite{lou2022testing}.

We argue two critical challenges exist in the field of testing MSF-based perception systems. Firstly, compared with the testing of single-sensor perception systems, testing MSF-based perception systems would require the synthesis of modality-consistent tests across different sensors (e.g., a car appears in an image should have the same pose in the corresponding point cloud). Therefore, \reviseline{a} mere combination of image-based and point cloud-based testing methods cannot guarantee such consistency. To address this challenge, Gao~et al. collected a set of perturbations that can be leveraged for testing MSF systems~\cite{gao2023benchmarking}. However, the testing efficiency is highly limited due to a lack of proper guidance when generating test cases. Another challenge comes from the realism of synthesized test cases. 
For perception systems operating in the physical world, the realism of test cases directly impacts the value of detected errors. Some existing perturbations used in testing, such as adversarial noises, seldom occur in real-world environments. Moreover, simple perturbations also limit the diversity of test generation from different perspectives (changing the number of cars and pedestrians, etc.). To address this, Wang et al. proposed an object insertion method for testing image-based object detectors~\cite{wang2020metamorphic}. However, without appropriate physical constraints, the inserted object might appear in invalid positions (e.g., a car hanging in the air) or result in incorrect perspective relationships (e.g., an occluded but visible building). 
To address these issues, we propose {\tool}, an automated testing method for MSF-based perception systems based on physical-aware multi-modal object insertion. Given a multi-modal test seed (i.e., a pair of image and point cloud frames), {\tool} automatically selects a 3D object instance (e.g., a car) from database then inserts it into the original data. To generate semantically plausible test data (i.e., the newly inserted car is on the road with correct heading), {\tool} first searches for valid poses of the inserted object. Then, {\tool} synthesizes realistic images and point cloud frames to handle the occlusion between the inserted object and the background data. Finally, {\tool} leverages a fitness-guided approach to insert the object for the purpose of synthesizing more challenging test cases for systems under test. {\tool} further employs metamorphic relations between the synthesized data and seed data to automatically detect faulty perception results.
{\tool} can generate modality-consistent and realistic tests from the seed testing data with physical-aware virtual sensors.

To evaluate the effectiveness and efficiency of {\tool}, we conduct experiments on three popular MSF-based object detection systems. We further experiment {\tool} with two single-sensor object detection systems (camera-only and LiDAR-only) to evaluate the generalization ability of {\tool}.
We find that {\tool} is capable of generating realistic and modality-consistent test data to satisfy test input specifications for both single-sensor and MSF-based perception systems.
The experimental results also demonstrate that {\tool} can effectively detect hundreds of erroneous behavior across different MSF-based perception systems.
We further retrain the selected perception systems with test data generated by {\tool}, finding that the system performance (measured by average precision) can be improved by 24\% on average.

In summary, this work makes the following contributions:

\begin{itemize}[leftmargin=*]
    \item \textbf{Method.} We propose an automated testing approach for MSF perception systems based on fitness-guided metamorphic testing. 
    Specifically, we leverage a physical-aware approach to insert new object instances to critical positions of the test seed to generate realistic and modality-consistent data.
    \item \sloppy \textbf{Tool.} We implement the above method into a tool {\tool}. 
    To the best of our knowledge, {\tool} is the first automated testing tool for MSF-based perception systems. 
    We have released {\tool} and the multi-modal data generated by {\tool} on our anonymous supplementary website: \href{https://sites.google.com/view/msftest}{https://sites.google.com/view/msftest}.
    \item \textbf{Evaluation.} We conduct extensive experiments to evaluate the performance of {\tool} with five perception systems. The results show that {\tool} can generate realistic and modality-consistent test data and efficiently detect erroneous system behavior. Retraining an MSF system with the generated test cases can significantly increase its robustness.
\end{itemize}

%







\begin{figure}[t]
    \centering
    \includegraphics[width = 0.9\linewidth]{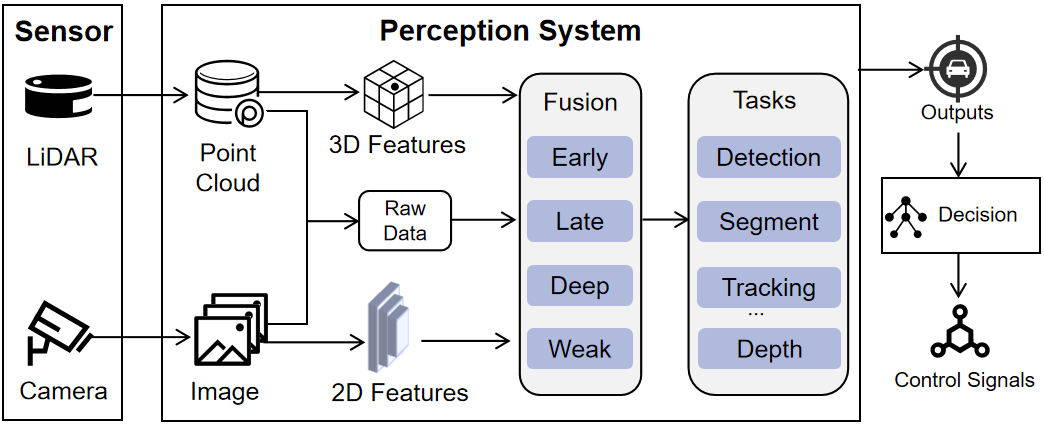}
    \caption{The workflow of MSF perception systems.}
    \label{fig: msf_example}
\end{figure}

\section{Background}

\subsection{Multi-Sensor Fusion Perception Systems}
\label{bg: ps}

Perception systems rely on sensors to capture real-world information in a particular data format. Subsequently, they process and fuse data from various different sensors using specific algorithms to help an intelligent system understand the external environment~\cite{survey_huang2022multi}.
As shown in Fig.~\ref{fig: msf_example}, perception systems are commonly integrated as a module within intelligent software systems (e.g., a self-driving car, a robotic, or an unmanned aerial vehicle). They are responsible for executing specific perception tasks, such as object detection and object tracking. The perception results are then communicated to downstream decision modules, such as planning and control, to facilitate the smooth operation of the entire intelligent system.
As a result, the quality and reliability of the entire system are significantly influenced by its perception systems.

To further enhance the performance of the system, most industrial-level systems~\cite{peng2020apollo} employ the multi-sensor fusion (MSF) strategy to avoid inherent perception limitations of individual sensors, leading to improved accuracy and reliable sensing capabilities.
For example, a camera can capture rich semantic information but is sensitive to the environmental changes (e.g., rain and snow), while LiDAR can provide high-quality 3D geometric information but lacks the capability of semantic understanding.
The camera-LiDAR fusion is also highly favored among different multi-sensor configurations.

To better fuse heterogeneous data with different characteristics and storage structure, researchers further propose AI-enabled MSF techniques.
Based on the stage of data fusion, these fusion techniques can be categorized into four different mechanisms at a high level: \textit{early fusion}, \textit{late fusion}, \textit{deep fusion}, and \textit{weak fusion}~\cite{survey_feng2020deep}.
\textbf{Early fusion} is the fusion of raw or pre-processed sensor data, which usually fuses heterogeneous data by converting them into the same data type according to rules.
While this fusion mechanism has low computational and memory requirements, its inflexibility and significant limitations have led to its infrequent standalone use. 
\textbf{Late fusion} directly merges the output results from both the LiDAR and the camera branch to make the final prediction.
Each branch independently processes data from sensors without relying on specific network architecture, which makes it more flexible and efficient.
\textbf{Deep fusion}, on the other hand, combines hidden features from different branches at varying depths to gain rich semantic information.
Frequent interactions between different branches enable the network to learn cross-modalities with diverse feature representations.
\textbf{Weak fusion} commonly uses rule-based methods to transform data from an additional branch to provide guidance for processing data in the main branch. 
A typical example of weak fusion is extracting the frustums in the point cloud data using the 2D detection bounding boxes from the image as guidance.

\subsection{The \reviseline{Preliminaries} of Object Detection}
\label{bk: object_detecion}
Object detection is a fundamental perception task for intelligent machines, enabling them to understand external environments. The primary objective of object detection is to estimate bounding boxes around objects of interest and predict their classification labels. This task can be categorized into two types based on dimensionality: 2D and 3D object detection.
2D object detection is typically performed on image data. It involves estimating a bounding box for each object, providing the object's position as $[x, y]$ and the bounding box's width $w$ and height $h$ in pixels.
In contrast, 3D object detection goes beyond 2D by additionally estimating the position $[x, y, z]$ of each object in 3D space, along with the bounding box's length $l$, width $w$, height $h$, and orientation angles $[roll, pitch, yaw]$. It is worth noting that, in the context of autonomous driving, practitioners usually consider only the $yaw$ orientation (i.e., the heading) for simplicity.
3D object detection can be performed on image data, point cloud data, or a fusion of both, depending on the available sensors and the specific application requirements.
Given different input data types (i.e., images or point clouds), object detection systems may employ diverse model architectures and pipelines. For image data, popular object detection systems utilize convolutional neural networks (CNNs) as the backbone to gather rich semantic hidden features~\cite{redmon2018yolov3,liu2016ssd,ren2015faster}. Subsequently, a head network is used to predict the bounding box's position, size, and the object's classification label.
Different from image data, point cloud data includes a set of orderless 3D points. Representative point cloud-based detection systems leverage two strategies to obtain semantic features from the point cloud: (1) voxel-based and (2) point-based methods. Voxel-based methods partition the point cloud into several fixed-resolution 3D grids (voxels) and use 3D CNNs as the backbone to extract point cloud features~\cite{yan2018second,zhou2018voxelnet}. By contrast, point-based methods directly extract features from raw point clouds via fully-connected networks~\cite{qi2017pointnet,qi2017pointnet++} or specialized convolution operations~\cite{liu2020closerlook3d} for points for 3D detection.
MSF-based object detection systems further take both these two input data types into account. Given different fusion strategies, an MSF-based object detection system might either extract features from each sensor's perception~\cite{huang2020epnet} or project the information captured in one modality to another modality before extracting features~\cite{chen2017multi}.

The accuracy of object detection is usually measured by IOU (intersection over union)~\cite{iou_padilla2020survey} and AP (average precision)~\cite{ap_everingham2010pascal}.
IOU measures the overlap area between a ground-truth bounding box $B_{g}$ and a predicted bounding box $B_{p}$ over their union. The computation of IOU can be represented as {\small $IOU={\operatorname{area}\left(B_{p} \cap B_{g}\right)}/{\operatorname{area}\left(B_{p} \cup B_{g}\right)}$}.
The object is \textit{successfully detected} in case IOU is greater than a given threshold $\tau$. In this paper, we set the IOU threshold $\tau$ as 0.5, \reviseline{which is consistent with the previous study~\cite{wang2020metamorphic}}.
AP is used to measure the performance of the overall detection performance on a dataset, which can be derived by computing the area under the Precision/Recall curve as 
\begin{equation}
\left.\mathrm{AP}\right|_{R}=\frac{1}{|R|} \sum_{r \in R} \rho_{\text {interp }}(r)
\label{eq: ap}
\end{equation}
where $\rho_{\text {interp }}$ is the interpolation function, which is defined as: 
$\rho_{\text {interp }}(r)=\max _{r^{\prime}: r^{\prime} \geq r} \geq\left(r^{\prime}\right)$, 
where $\rho(r)$ gives the precision at recall level $r$. 
In this paper, we apply forty equally spaced recall levels recommended by KITTI, i.e., $R_{40}=$ $\{1 / 40,\;2 / 40,\;\ldots,\;1\}$.

For a more in-depth evaluation of the object detection tasks, Wang et al.~\cite{wang2020metamorphic} classified object detection errors into three categories, i.e., \textit{Recognition failures}, \textit{Localization failures}, and \textit{Classification failures}. Note that in this paper, we mainly focus on recognition failures and localization failures.

Recognition failures include two independent types of errors, i.e. \textit{Object missing} and \textit{False detection}. 
Object missing refers to the case that an object detection system fails to recognize an existing object, while 
false detection refers to the case that the system treats an arbitrary region without objects as an ``object''.
Given a ground-truth bounding box $B_{g} \in \mathbb{GT}$ and a predicted bounding box $B_{p} \in \mathbb{DT}$, the object missing can be formalized as: 
\begin{equation}
\exists B_{g} \in \mathbb{GT} \wedge \forall B_{p} \in \mathbb{DT},  IOU(B_{g},B_{p}) \leq 0
\end{equation}
and the false detection can be be formalized as:
\begin{equation}
\forall B_{g} \in \mathbb{GT} \wedge \exists B_{p} \in \mathbb{DT} ,  IOU(B_{g},B_{p}) \leq 0 
\end{equation}
Localization failures refer to the errors that an estimated bounding box is too large or too small, which can be formalized as:
\begin{equation}
\exists B_{g} \in \mathbb{GT}  \wedge \exists B_{p} \in \mathbb{DT}_{l} , 0 <IOU(B_{g},B_{p}) \leq \tau
\end{equation}
where $\mathbb{DT}_{l}$ is detection bounding boxes with neither successful detection nor false detection. 

\section{Approach}
\label{ap: sensor}

\begin{figure*}[t]
    \centering
    \includegraphics[width = 0.95\linewidth]{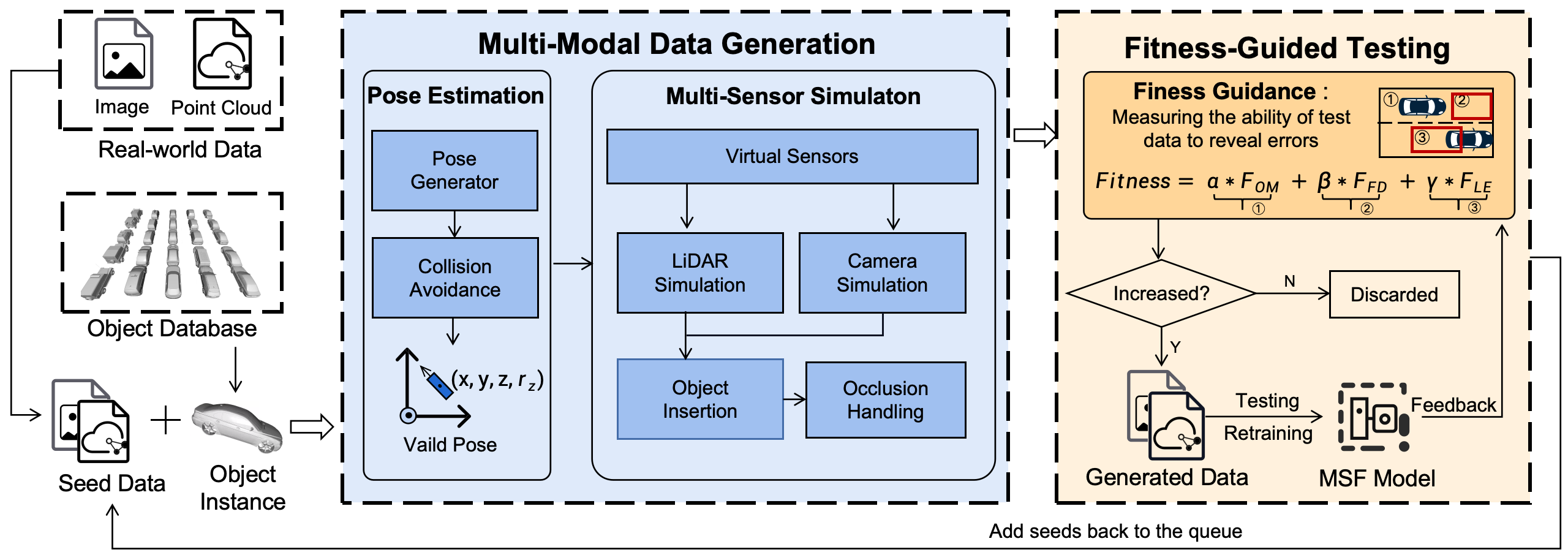}
    \caption{The workflow of {\tool}.}
    \label{fig: workflow}
\end{figure*}

In this section, we introduce the design of {\tool}. 
Fig.~\ref{fig: workflow} presents the high-level workflow of {\tool}.
Given a background multi-modal data recorded from real-world and an object instance selected from the object database, 
{\tool} first executes the \textit{pose estimation} module to calculate the valid locations and orientations of an object to be inserted. 
Then the \textit{multi-sensor simulation} module renders the object instance in the form of both image and point cloud given the calculated poses in a physical-aware virtual simulator. The \textit{multi-sensor simulation} module further merges the synthesized image and point cloud of the inserted object with the background data and carefully handles the occlusion. 
These two modules form the {\tool}'s \textbf{multi-modal test data generation} pipeline.
Finally, the realistic multi-modal test data can be efficiently generated through \textbf{fitness guided metamorphic testing}. We detail each module of {\tool} in the following.

\subsection{Pose Estimation}
\begin{figure}
    \centering
    \subfigure[Invalid insertion with plane-equation based sampling]{
        \label{fig: incorrect_position}
        \includegraphics[width=0.4\linewidth]{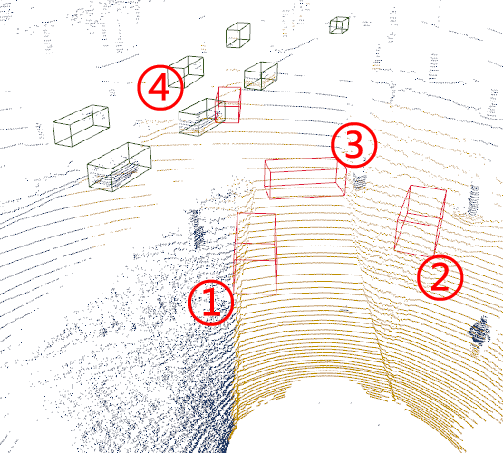}
    }
    \hspace{6mm}
    \subfigure[Valid insertion with {\tool}]{
        \label{fig: valid_pose}
        \includegraphics[width=0.4\linewidth]{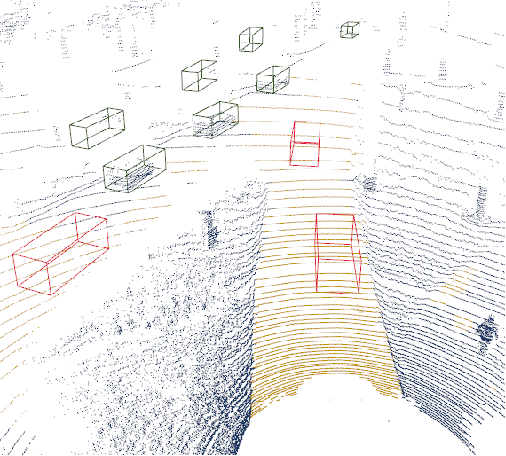}
    }
    \vfill
     \subfigure[Zoom-in views of different invalid insertion]{
        \label{fig: pose_eg}
        \includegraphics[width=0.2\linewidth]{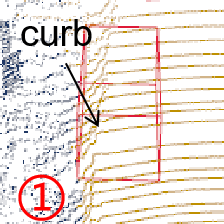}
        \hspace{2mm}
        \includegraphics[width=0.2\linewidth]{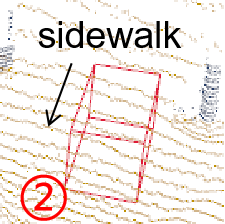}
        \hspace{2mm}
        \includegraphics[width=0.2\linewidth]{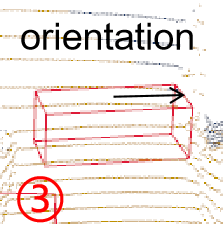}
        \hspace{2mm}
        \includegraphics[width=0.2\linewidth]{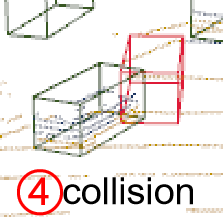}
    }

    \caption{Validity of object insertion with different pose estimation methods.}
    \label{fig:invalid_pose}
\end{figure}

Given the background data and an object instance, the pose estimation module aims to estimate possible valid locations and orientations to create a plausible scene in the real world after inserting the object. This is critical for synthesizing realistic test data for real-world perception systems because failing to address it could result in semantically invalid data. For instance, a synthesized test case should be considered invalid if a car is inserted outside of the road or if the car's heading is incorrect (Fig.~\ref{fig: pose_eg}).



\vspace{1mm}
{\noindent \bf Pose Generator.}
Given a candidate scene $m$ (a.k.a. a test seed, composed by a pair of image and point cloud $\langle image, pc \rangle$) to insert and an object $o$ to be inserted, {\tool}'s pose generator first samples a set of possible positions and orientations $\mathcal{P}$ ($pose\_i \in \mathcal{P}$, $pose\_i=[x_i, y_i, z_i, r_{zi}]$ is in the LiDAR's coordinate) in the corresponding 3D space. To obtain $\mathcal{P}$, one critical challenge is to split road planes from the 3D scene. 
State-of-the-art (SOTA) methods~\cite{yan2018second, lin2022taulim} use plane segmentation algorithms (e.g., fitting plane equations with RANSAC~\cite{fischler1981random}) to obtain road planes. However, we posit that the road planes obtained through fitting plane equations may inadvertently include non-road objects such as sidewalks, median strips, etc.
Therefore, we utilize CENet~\cite{cheng2022cenet}, a concise and efficient point cloud semantic segmentation model, to split the road point cloud from the background point cloud.
Subsequently, we meshify the road point cloud to reconstruct the road surface and obtain sample insertion positions and orientations.
As shown in Fig.~\ref{fig:invalid_pose}, compared to plane-equation based sampling method that \reviseline{is} used in the existing testing tool~\cite{lin2022taulim}, {\tool} can accurately generate possible positions for object insertion, even in complicated road conditions such as intersections. 



{\noindent \bf Collision Avoidance.}
To avoid collision with existing objects from the background scene $m$, we first calculate the minimal 3D bounding box $B_{o}$ of the object instance to be inserted. 
Then, given a possible pose of $o$, $pose\_i$, we check if the 3D box $B_{o}$ contains any other objects from the background point clouds. If the IOU between $B_{o}$ and the background point cloud is greater than 0, we consider $pose\_i$ as an invalid pose since it leads to a collision with the existing objects.



\subsection{Physical-Aware Multi-Sensor Simulation}
One significant threat to the realism of test cases generated by the existing methods is that the random insertion might violate the basic laws of the physical world.
For example, images captured by a camera follow the basic law that objects farther  away are smaller. Similarly, point clouds recorded by LiDAR should become sparser when the inserted object is farther away.
Moreover, different sensor configurations could result in varying field of views (FOVs) and sensing results. Therefore, when rendering an object for insertion, it is critical to use the same sensor settings (camera resolution, number of LiDAR beams, calibration files, etc.) as those used for the background scene.
Consequently, {\tool} employs a physical-aware sensor simulation module to render images and point clouds for the object to be inserted.
We first construct a set of virtual sensors to simulate their operating modes in the physical world to generate modality-consistent data given their configuration parameters. Then, we carefully handle the occlusion when inserting the object $o$ into the candidate scene $m$.

\begin{figure}[t]
\centering   
\subfigure[View along the y-axis]{\label{fig:b}\includegraphics[width=0.4\linewidth]{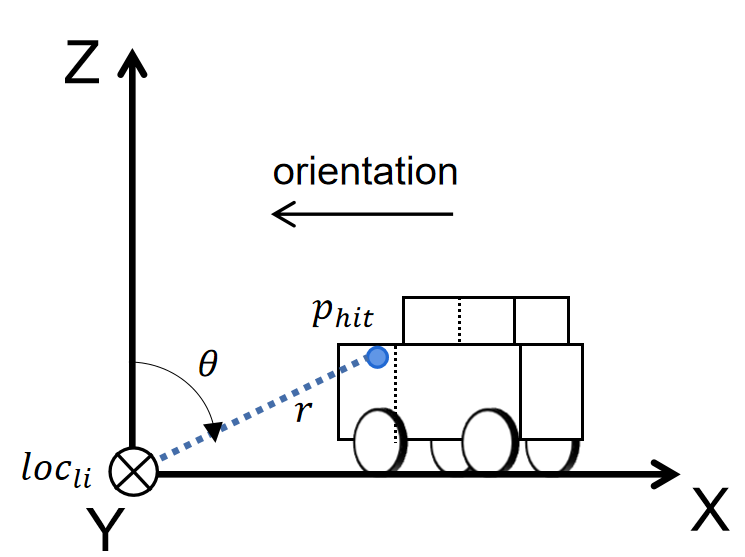}}
\label{fig: sensor_along_y}
\hspace{10mm}
\subfigure[Top view]{\label{fig:a}\includegraphics[width=0.4\linewidth]{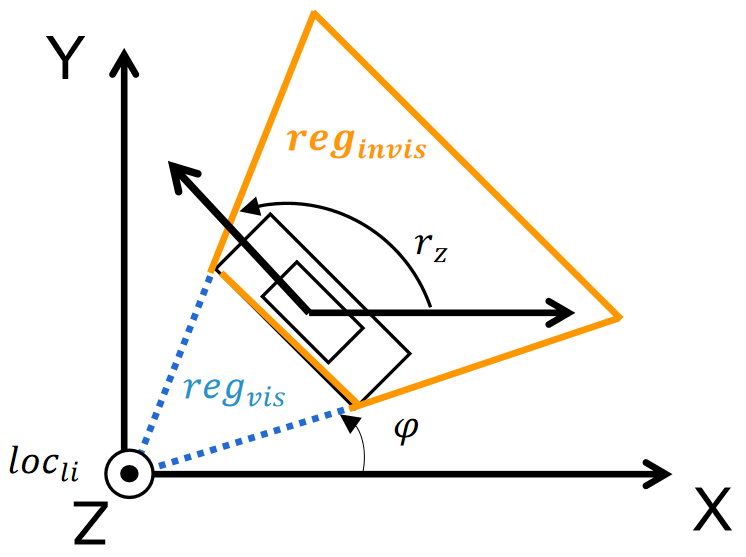}}
\label{fig: sensor_along_z}
\caption{Virtual LiDAR emitting lasers to hit a car. (a) View along the y-axis, (b) top view.}
\label{fig: sensor}
\end{figure}

\vspace{1mm}
{\noindent \bf LiDAR Simulation.}
To simulate laser beams of a LiDAR sensor, given the LiDAR horizontal field of view, vertical field of view, the number of beams and angular resolution, we first create a set of rays centered on the virtual LIDAR scanner to simulate each laser beam emitted. 
Then we leverage the \textit{ray casting} proved by Open3D~\cite{zhou2018open3d} to simulate laser emission on the object $o$. As shown in Fig.~\ref{fig: sensor}, the ray casting calculates the 3D coordinates of the first hit point $p_{hit}(r,\theta,\phi)$ of the ray on the target surface given the starting point $loc\_lidar$ and direction of the ray, where $r$ is the distance, $\theta$ is the polar angle and $\phi$ is the azimuthal angle in spherical coordinates.
Whenever an obstacle is inserted into the scene, we execute the ray casting to obtain the updated coordinates in the point cloud.

Considering that the point cloud data captured by a real-world LiDAR could include inevitable noises, we randomly remove a small portion of the point to simulate the laser which is not to be detected by the receiver due to insufficient intensity~\cite{kashani2015review}. We further add Gaussian noise to simulate the measurement noise of the sensor~\cite{fang2021lidar}. Note that the proposed virtual LiDAR can simulate any type of laser scanner with the given LiDAR configurations, including 
its extrinsic calibration, FOVs and resolutions, the number of beams, etc. We refer readers to our supplementary website for more details.

\vspace{1mm}
{\noindent \bf Camera Simulation.}
We leverage Python API provided by Blender, an open-source 3D computer graphics software, to build our virtual camera sensor.
We first build a virtual camera given its configurations, including intrinsic and extrinsic calibration, lens length, resolution, etc. 
Then, we calculate the pose of the object $o$ in the camera coordinates system $pos\_cam = \mathbf{T}_{\text {velo }}^{\text {cam }} pos\_li$ based on the transformation matrix $\mathbf{T}_{\text {velo }}^{\text {cam }}$ and its corresponding position $pose\_i$ in the LiDAR coordinate.
Finally, we place the object $o$ in the position $pos\_cam$ and render an image with our virtual camera.
Note that we use the minimum 2D rectangle box of the image object $o$ as the ground truth bounding box for 2d object detection.
To further improve the realism of the synthesized test cases, we leverage a model S$^{2}$CRNet~\cite{liang2022spatial} \reviseline{with pre-trained weights released by the authors~\footnote{\href{https://github.com/vinthony/S2CRNet-demos}{https://github.com/vinthony/S2CRNet-demos}}} to naturally blend the object $o$ into the target scene $m$ by adjusting the color of the inserted object. Therefore, the inserted object could have near consistent illumination and color balancing compared with the target scene.

\vspace{1mm}
{\noindent \bf Occlusion Handling.}
When inserting an object, occlusion should be carefully addressed. For instance, the inserted object should not block any objects that are closer to the sensor according to the distance to the sensor. 
To handle the point cloud occlusion, we remove the point clouds that are occluded by the inserted objects from the 3D scene according to their geometric relationships. 
As shown in Fig.~\ref{fig: sensor} (b), the ray emitted from the virtual laser scanner intersects the object and divides the space into the visible region $reg\_vis$ and the invisible region $reg\_invis$. 
We sequentially infer the occlusion relation and remove the points in $reg\_vis$ for each object according to their distance to the virtual LiDAR.
To handle the image occlusion, we record the distance from the center of each object to the virtual camera and ensure the inserted object $o$ only blocks the pixels that are farther than $o$ in the 3D space. \reviseline{To avoid false positives on original objects that are occluded by the inserted objects, we calculate an occlusion ratio for each existing object after each round of insertion. If the occlusion ratio of an occluded object is higher than 90\%, we exclude this object from the evaluation by labeling it as ``DontCare.''}

\subsection{Metamorphic Relations}

Manual labeling of our generated test cases could be time-consuming and labor-extensive.
To address this, we leverage metamorphic relations (MRs)~\cite{chen2018metamorphic} to create test oracles. MRs describes the necessary properties of a target software in terms of inputs and their expected outputs~\cite{zhou2004metamorphic}. The violation of MRs often indicates potential defects. 


{\tool} is designed for MSF perception systems based on metamorphic testing.
Specifically, we denote the MSF perception system as $PS$ that detect the results with multi-model data $m \in \mathbb{M}$ 
including 2D image and 3D point cloud.
Given a set of object instances $\mathbb{O}$, 
an MR can be formalized as follows:
\begin{equation}
    MR_1:~\forall o \in \mathbb{O} \wedge \forall m \in \mathbb{M}, \zeta\{PS \llbracket m \rrbracket \cup GT_{o}, PS \llbracket \sigma(m,o) \rrbracket\}
\end{equation}
where $\sigma$ is the insertion operator for inserting an object instance $o$ into the background scene $m$, $GT_o$ is the ground truth of the object $o$ (i.e., the estimated bounding box in object detection task) and $\zeta$ is a criterion asserting the equality of $PS$ results.

$MR_1$ is built upon the following two facts: (1) the object insertion operator should not change the correct prediction of $PS$;
(2) and the inserted object should be detected correctly.
However, asserting the equality of $PS$ outputs is too strict and thus can lead to a large number of false positives due to a slight drift of the detection results. 
Therefore, we follow the previous work~\cite{guo2022lirtest, wang2020metamorphic} to leverage soft equality criteria $\zeta$ derived from Average Precision (AP).
Given the MR, we can simply obtain the test oracle information without manual annotation by checking if the MR is violated. 

\subsection{Fitness-Guided Testing}
To boost the testing efficiency of {\tool}, we propose a fitness-guided testing process for object insertion. Thus, we can keep a test case $m'$ with high fault-revealing capability from the seed test $m$. 


\vspace{1mm}
{\noindent \bf Fitness Metric.} 
We design a fitness metric that measures the likelihood of a test data to reveal errors.
Our fitness metric consists of three fault-revealing capability scores, i.e., object missing, fault detection, and location error (introduced in Sec.~\ref{bk: object_detecion}).

Given a ground-truth bounding box $B_{g} \in \mathbb{GT}$ and a predicted bounding box $B_{p} \in \mathbb{DT}$,
the object missing score can be expressed as follows:
\begin{equation}
F_{OM}= \sum_{B_{g} \in \mathbb{GT}} I_{OM}(B_{g}, B_{p}) *  \left(1 - \frac{\min(dis(B_{g}),dis\_max) }{dis\_max}\right)
\end{equation}
where $dis$ calculates the distance between a bounding box $B$ and the LiDAR position, $dis\_max$ is the max recognition distance of LiDAR, 
and $I_{OM}(\cdot)$ is an indicator function equal to 1 if and only if it is an object missing error.
The intuition behind the score is that the close object missing failure could lead to serious consequences, such as collisions.

The fault detection score can be expressed as follows:
\begin{equation}
F_{FD}= \sum_{B_{p} \in \mathbb{DT}} I_{FD}(B_{g}, B_{p}) * \left(1 - \frac{\min(dis(B_{p}),dis\_max) }{dis\_max}\right) * prob_{p} 
\end{equation}
where $prob_{p} $ is the confidence probability of the detection, and $I_{FD}(\cdot)$ is an indicator function equal to 1 if and only if it is a false detection error.
Similarly, a close and high-confidence fault detection might lead to wrong control decisions, e.g., emergency braking.

Then, we compute location error scores if and only if it is neither missing detection nor fault detection. Let $\mathbb{GT}_{r} = \{B_{g}|I_{OM}(B_{g})=0 \wedge B_{g} \in \mathbb{GT}\}$  denotes the cases that do not contain object missing errors and $\mathbb{DT}_{r} = \{B_{p}|I_{FD}(B_{p})=0 \wedge B_{p} \in \mathbb{DT}\}$ denotes the cases that do not contain fault detection errors:
\begin{equation}
F_{LE}= \max_{B_{g} \in \mathbb{GT}_{r} \wedge B_{p} \in \mathbb{DT}_{r}} 1 - IOU(B_p,B_g)
\end{equation}
This score indicates that a larger difference between $B_p$ and $B_g$ could reveal a more serious detection error.

Finally, our fitness metric can be expressed as a weighted sum of the three scores:
\begin{equation}
\label{eq: fitness}
Fitness(m) = \alpha * F_{OM} + \beta * F_{FD} + \gamma *F_{LE}
\end{equation}
where $\alpha + \beta+ \gamma = 1$. 
Similar to the traditional software testing approach to increase the code coverage rate, {\tool} attempts to generate a test set that can increase this fitness score. 

{
\setlength{\dbltextfloatsep}{0pt}
\begin{algorithm}[t]  
\small
\caption{Fitness-guided testing of {\tool}.}  
\label{alg: guide} 
\LinesNumbered
    \KwIn{The test MSF system $PS$, 
    the set of seed multi-model data $\mathbb{M}$,
    the object database $\mathbb{O}$, 
    the maximum of object insertion $N$,
    the maximum number of trials $TRY\_NUM$.}
    \KwOut{the set of generated data $\mathbb{T}$, the set of ground-truth $\mathbb{G}$.}
    $\mathbb{T} \leftarrow \emptyset$\;
    \For{$m$ in $\mathbb{M}$}{
        $L,\;succFlag\leftarrow LoadLabel(m),\;false$\; 
        $finess\_init \leftarrow Fitness(m)$\;
        \For{$i = 1,2,\ldots,N $}{
            Randomly sample an object instance $o$ from $\mathbb{O}$\;
            \For{$i = 1,2,\ldots,TRY\_NUM $}{
                 $pose \leftarrow  PoseGenerator(o,m)$\Comment{pose estimation}\;
                \If{$CollisionAvoidance(pose,m)$}{
                    \textbf{continue};
                }
                $pc \leftarrow LidarSimulator(o,m)$\Comment{sensor simulation}\;
                $image \leftarrow CameraSimulator(o,m)$\;
                $m' \leftarrow Insertion(pose,pc,image,m)$\;
                $m' \leftarrow OcclusionHandler(m')$\;
                ${lb'} \leftarrow  LabelGenerator(o,pose_{o},m')$\;
                \tcp{save test $m'$ with higher fitness}
                $finess\_score \leftarrow Fitness(m')$\;
                \If{$finess\_score > finess\_init$}{
                    $succFlag,\;m \leftarrow True,\;m'$\;
                    $L \cup \{lb'\}$\;
                    $finess\_init \leftarrow finess\_score $\;
                    \textbf{break};
                }
            }
        }
    \If{$succFlag$}{
            $\mathbb{T},\;\mathbb{G}  \leftarrow \mathbb{T}\cup\{m'\},\;\mathbb{G}\cup\{L\}$\;
        }    
    }
    \textbf{Return: } $\{\mathbb{T},\mathbb{G}\}$ ; 
\end{algorithm}
}
{\noindent \bf Testing Workflow.} 
Algorithm~\ref{alg: guide} presents the testing workflow of {\tool} guided by our fitness metric. The algorithm takes 
an MSF perception system $PS$, 
a set of seed multi-model data $\mathbb{M}$,
an object database $\mathbb{O}$, 
and the constant maximum number of trials $TRY\_NUM$ 
and object insertion $N$ 
as input. The goal is to create a set of critical test cases $\mathbb{T}$ with corresponding ground-truth $\mathbb{G}$ for $PS$.
The algorithm first selects a valid pose of object instance from the \textit{pose generator} (Lines 8-10). Then, the sensor simulation module renders the object-level point cloud and image before inserting them into the background data to obtain synthetic data $m'$ (Lines 11-14). 
After insertion, we generate the label of test case $m'$ (Line 15) and calculate the fitness score by Eq.~\ref{eq: fitness} (Line 16).
If the fitness metric increases, the generated test is retained as a new background scene for the next iteration of object insertion (Lines 17-21).
By contrast, the generated data is directly discarded, which indicates that $PS$ might not be prone to erroneous behavior in this case.
Finally, a successfully generated test case $m'$ with the label $L$ generated by $N$ iterations of object insertion is added to $\mathbb{T}$ and $\mathbb{G}$, respectively.

\section{Experimental Setup}

In this section, we introduce our experimental setup, including our implementation details, perception systems under test, datasets, and the research questions we investigate.

\subsection{Implementation Details}

In all experiments, we set the maximum number of trails $TRY\_NUM$ as 5 and the maximum number of insertions $N$ as 3 in Alg.~\ref{alg: guide}.
$\alpha, \beta, \gamma$ in Eq.~\ref{eq: fitness} are set to $0.5, 0.25, 0.25$, respectively.
We leverage the configurations of Velodyne HDL-64E lidar and PointGray Flea2 color camera provided by KITTI~\cite{kitti_geiger2012we} to build our virtual simulator in Sec.~\ref{ap: sensor}.
To conduct the experiments, we implement the {\tool} upon PyTorch 1.8 and Python 3.7. All experiments are conducted on a server with an Intel i7-10700K CPU (3.80 GHz), 48 GB RAM, and an NVIDIA RTX 3070 GPU (8 GB VRAM). 

\subsection{Perception Systems Under Test}

To evaluate the effectiveness and efficiency of {\tool}, we choose three SOTA camera-LiDAR fusion perception systems covering different fusion mechanisms from an MSF benchmark~\cite{gao2023benchmarking} as the SUTs (system under test). Additionally, we consider one single-sensor detection system for both camera and LiDAR in our experiment. We compare five perception systems and their original performance (AP) on KITTI 2D/3D detection benchmark in Table~\ref{tab:SUTs}.

\begin{table}[t]
    \footnotesize
    \renewcommand\arraystretch{1.2}
    \setlength{\tabcolsep}{7.5pt}
    \centering
    \caption{Perception systems under test and their performance on KITTI benchmark.}
    \begin{tabular}{lcccc}
        \toprule
        \textbf{SUT} & \textbf{Fusion} & \textbf{Year} & \textbf{2D Det. AP} & \textbf{3D Det. AP}\\
        \midrule
         {CLOCs}~\cite{pang2020clocs} & Late fusion & 2020 & 89.82 & 89.37\\
         {EPNet}~\cite{huang2020epnet} & Deep fusion  & 2020 & 89.84 & 89.54\\
         {FConv}~\cite{wang2019frustum} & Weak fusion & 2019 & 90.19 & 90.42\\
         {$\star$ Second}~\cite{yan2018second} & ---  & 2018 & --- & 89.09\\
         {$\star$ CasRCNN}~\cite{cai2018cascade} & --- & 2018 & 90.23 & ---\\
         \bottomrule
    \end{tabular}
    \label{tab:SUTs}
\end{table}


\subsection{Dataset}

We select our test seeds from KITTI dataset~\cite{kitti_geiger2012we}. KITTI is one of the most popular autonomous driving datasets, which provides images and point clouds, with detailed sensor configurations. 
KITTI's data are collected by four high-resolution cameras, a Velodyne HDL-64E LiDAR and an advanced positioning system from multiple categories real-world driving scenarios, such as cities, residential areas, and roads.
The KITTI object detection dataset contains 7481 pairs of image and point cloud data with ground-truth labels.
The ground-truth labels include both annotated 2D and 3D boxes, difficulty levels and category labels for objects of interest.
In this paper, we focus on the detection of car objects at the moderate difficulty level.

For our object databse $\mathbb{O}$, we utilize ShapeNet~\cite{chang2015shapenet}. ShapeNet is a richly-annotated and large-scale dataset of 3D object models.
It contains over 220,000 object-level models from different categories.
In this paper, we build a multi-modal object database from the \textit{car} category in ShapeNet, which consists of 3483 objects. We further filtered out damaged models (e.g., empty model files) and uncommon vehicles on the road (e.g., a racing car). 
Consequently, our object database includes 1674 vehicles from five categories: \textit{sedan}, \textit{coupe}, \textit{suv}, \textit{cab}, and other unclassified cars.

\subsection{Research Questions}
To evaluate {\tool}'s performance, we conduct both quantitative and qualitative experiments to answer the following three research questions (RQs):

\vspace{-1pt}
\begin{itemize}[leftmargin=*]
    \setlength\itemsep{0.5mm}
    \item \rqone
    \item \rqtwo
    \item \rqthree
\end{itemize}

To answer {\bf RQ1}, we verify the effectiveness of two modules in {\tool}'s multi-modal data synthesis pipeline: \textit{Pose Estimation} and \textit{Multi-sensor Simulation}. We replace each corresponding module with a baseline module to conduct control experiments. Specifically, we replace {\tool}'s \textit{Pose Estimation} module with a \textit{Random Pose} module and {\tool}'s \textit{Multi-sensor Simulation} module with \textit{Cut\&Paste} module. The \textit{Random Pose} module randomly selects a pose for the object $o$ without any constraints. The \textit{Cut\&Paste} module directly copies an object's image and point cloud from other data frames in the KITTI dataset without any re-rendering. 
Hence, we can obtain four multi-modal data synthesis pipelines. We denote them as (1) $C\&P$ + $Pose_{ran}$, (2) $C\&P$ + $Pose_{est}$, (3) $Sim$ + $Pose_{ran}$, and (4) $Sim$ + $Pose_{est}$ ({\tool}). We then conduct both quantitative and qualitative assessments to verify the data realism.

{\bf Quantitative Assessment.} We randomly select 200 data pairs from KITTI's validation dataset as the initial seeds for each pipeline and compared the realism of the data generated by each pipeline. Then we utilize \textbf{FID} (Frechet Inception Distance)~\cite{heusel2017gans} and \textbf{FRD} (Frechet Range Distance)~\cite{zyrianov2022learning} to measure the realism of the image and point cloud, respectively. \textbf{FID~\cite{heusel2017gans}} evaluates the squared Wasserstein distance between feature vectors extracted from the inception-v3~\cite{szegedy2016rethinking} model from the generated samples $G$ and real samples $R$. The computation of FID can be represented as:
\begin{equation}
    \mathrm{FID(R,G)}=\left\|\mu_r-\mu_g\right\|^2+\operatorname{Tr}\left(\Sigma_r+\Sigma_g-2\left(\Sigma_r \Sigma_g\right)^{1 / 2}\right)
\end{equation}
where $\mu$ and $\Sigma$ represent the mean values and covariance matrix of generated samples respectively and $\operatorname{Tr}$ denotes the trace operator of the matrix.
Similar to FID, \textbf{FRD~\cite{zyrianov2022learning}} measures the 
similarity between two sets of point cloud data using the feature vectors from a pre-trained RangeNet++~\cite{milioto2019rangenet++}.
Note that we set the configurable parameters of these metrics with default/recommended settings.

We further propose a new \textbf{Modality-Consistency (MC)} metric, which is specifically designed for our multi-modal object insertion.
For each inserted object ${o \in O}$, we first calculate its minimum bounding box in the point cloud and project it onto the image as $B^p_{g}$. Then, we measure the average IOU between 2D bounding boxes of an image $B^i_{g}$ and projected bounding boxes $B^p_{g}$ through:
\begin{equation}
MC = \frac{1}{|O|} * \sum_{o \in O}(IOU(B^i_{g},B^p_{g})) 
\end{equation}
Intuitively, this metric measures if the inserted object has consistent poses and dimensions in both image and point cloud data.

\reviseline{To better demonstrate the realism of test data generated by different pipelines, we also implement two SOTA object-insertion based testing methods for single-sensor perception systems (i.e., MetaOD~\cite{wang2020metamorphic} (camera-based) and TauLim~\cite{lin2022taulim} (LiDAR-based)) and leverage their combination as another comparison baseline for our quantitative assessment in this RQ.}

{\bf Qualitative Assessment.} We further conduct a user study to qualitatively assess the naturalness of {\tool}'s generated multi-modal data. We recruited sixteen participants through the mailing list of the CS department at a research university. All participants had a minimum of a master's degree in SE/CS. Seven out of sixteen participants had more than two years of experience in the field of autonomous driving. 
During the study, we randomly selected twenty data instances as test seeds. We then ask a participant to rank the multi-modal data synthesized by four different pipelines from each seed through a questionnaire.  For each of the twenty data instances, a participant needs to rank the data quality from three perspectives: (1) the image's naturalness, (2) the point cloud's naturalness, and (3) the modality-consistency between a pair of image and point cloud. To mitigate bias, we randomly assigned the order of data synthesized by different pipelines for each test seed. We further refer readers to our supplementary website for the complete questionnaire. After finishing the user study, we use the one-side Wilcoxon rank-sum test~\cite{cuzick1985wilcoxon} to verify if any of the four pipelines is significantly preferred by the participants.
To answer {\bf RQ2}, we utilize random testing as \reviseline{a} baseline to evaluate the effectiveness of our proposed fitness-guided testing strategy. 
\reviseline{We also include MetaOD~\cite{wang2020metamorphic} (camera-based), TauLim~\cite{lin2022taulim} (LiDAR-based), and their combinations as the comparison baselines.}
Specifically, we randomly select 200 data instances as the initial seeds from the validation set of the KITTI dataset to generate test cases using both random testing and {\tool}'s fitness-guided testing.
We then calculate the AP difference between the original datasets and the generated test cases of a perception system under test for each configuration. 
Furthermore, we count the number of errors with each error category, i.e., \textit{object missing}, \textit{false detection}, and \textit{location error} (see Sec~.\ref{bk: object_detecion}).
Note that we only count the errors when a perception system has a confidence level greater than 0.5 for the purpose of mitigating its trivial faults. Each experiment is repeated five times to mitigate the effect of randomness.
To answer {\bf RQ3}, we retrain all five perception systems on the corresponding test cases generated by {\tool} \reviseline{or other baselines in RQ2}. For each system, we randomly generate 1000 test cases for each perception system. Then we select 20\% of the test cases and add them to the original KITTI training set. We keep the training configuration consistent with each system's original settings and parameters. We set the number of epochs as 40 for retraining. After retraining, we compare the performance of each retrained perception system with its original counterpart on the remaining 80\% of generated test cases. We retrain each system for \reviseline{five} times to mitigate the effect of randomness and compare its average performance.




\section{Result Analysis}


\subsection{RQ1: Realism Validation}


\begin{table}[t]
\caption{Realism of test data generated by different pipelines.}
\label{tab: rq_real_sim}
\footnotesize
\renewcommand\arraystretch{1.2}
\setlength{\tabcolsep}{7.5pt}
\begin{tabular}{lccc}
\toprule
\multicolumn{1}{l}{\multirow{2}{*}{\textbf{Data Generation}}} & \multicolumn{1}{c}{\textbf{Image}} & \multicolumn{1}{c}{\textbf{Point Cloud}} & \multicolumn{1}{c}{\textbf{Consistency}} \\
\multicolumn{1}{c}{}                                              & \multicolumn{1}{c}{FID~$\downarrow$}                & \multicolumn{1}{c}{FRD~$\downarrow$}             & \multicolumn{1}{c}{MC$~\uparrow$} \\
\midrule
$C\&P$ + $Pose_{ran}$                                              & 106.13           & 148.24             & 0.31                  \\
$C\&P$ + $Pose_{est}$                                             & 107.62            & 115.62             & 0.30                   \\
$Sim$ + $Pose_{ran}$                                              & 91.34             & 111.14             & 0.83                  \\
\reviseref{}$MetaOD+TauLim$                                                                       & \reviseref{}65.57                                       & \reviseref{}133.61                                      & \reviseref{}0.01                                       \\
\cellcolor{lightgray}$Sim$ + $Pose_{est}$ ({\tool})                                       & \cellcolor{lightgray}86.89             & \cellcolor{lightgray}88.87              & \cellcolor{lightgray}0.82                  \\   
\bottomrule
\end{tabular}
\end{table}

\begin{figure}[t]
    \centering
    \includegraphics[width = 0.85\linewidth]{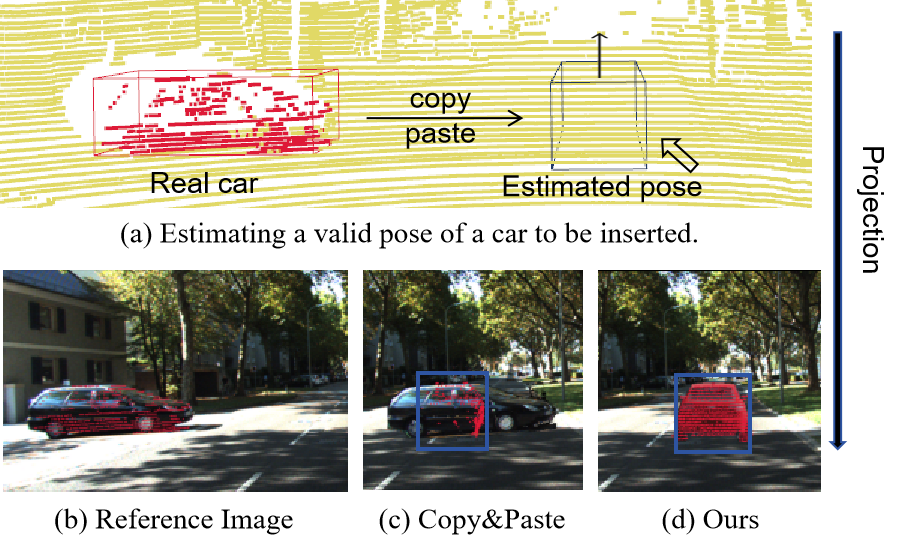}
    \caption{Modality consistency of data synthesized with different simulation methods.}
    \label{fig: mc}
\end{figure}

Table~\ref{tab: rq_real_sim} shows the quantitative results of different data synthesis pipelines. \reviseline{Compared with baselines, {\tool} achieves the second-lowest FID and the lowest FRD.} These results indicate that {\tool} can generate high quality and realistic image and point cloud data after the object insertion. \reviseline{We notice that MetaOD+TauLim achieves better image quality (FID). A plausible explanation is that MetaOD only inserts one object per image while {\tool} may insert multiple objects. Our further investigation shows that when only inserting one object, {\tool} can generate images with similar quality (FID: 62.53). However, the combination of MetaOD and TauLim produces the lowest MC, indicating that directly combining two single-sensor testing methods may produce data with high inconsistency.} We also find that compared with the \textit{Copy\&Paste} method, our \textit{Multi-sensor Simulation} module can generate much more modality-consistent data pairs of images and point clouds. A plausible explanation is that \textit{Copy\&Paste} does not take sensor's position and inserted object's pose into account. As the example shown in Fig.~\ref{fig: mc} (a), given an estimated pose to insert a car object, \textit{Copy\&Paste} directly copies a car object from existing data (Fig.~\ref{fig: mc} (b)). However, the image data can not be rotated to align with the given pose, resulting in modality inconsistency (Fig.~\ref{fig: mc} (c)) compared with {\tool} (Fig.~\ref{fig: mc} (d)). Furthermore, we find that our \textit{Pose Estimation} module outperforms the \textit{Random Pose} on the naturalness of the generated point cloud. By replacing \textit{Random Pose} with {\tool}'s \textit{Pose Estimation}, the FRD values decreased by 22\% and 21\% for \textit{Copy\&Paste} and \textit{Multi-sensor Simulation}, respectively. This is largely attributed to the fact that \textit{Random Pose} does not take physical constraints into account when inserting objects. Consequently, it inevitably inserts objects into invalid positions. 

\begin{figure}[t]
    \centering
    \includegraphics[width = 0.9\linewidth]{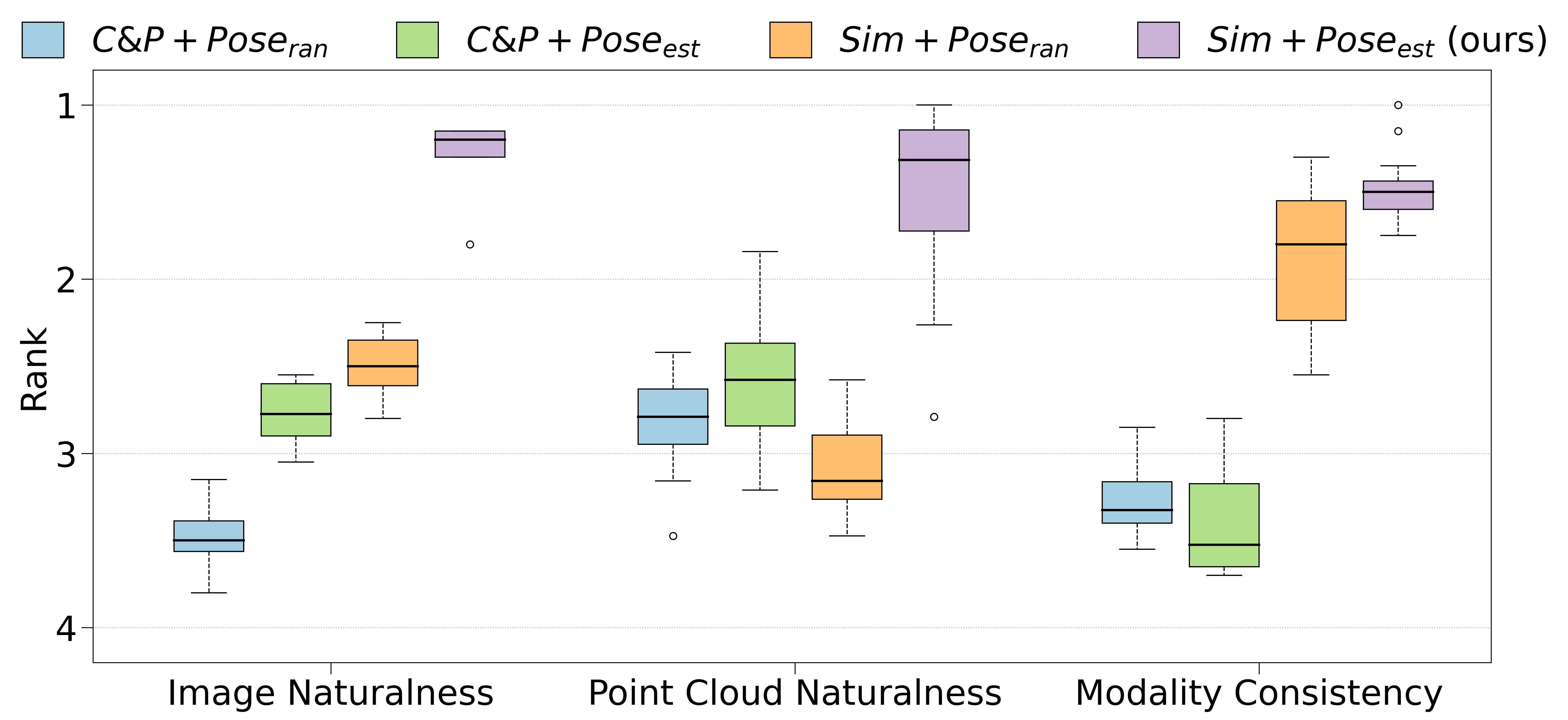}
    \caption{Participants' choices over different data synthesis pipelines in \textit{image naturalness}, \textit{point cloud naturalness}, and \textit{modality consistency}.}
    \label{fig: human_val}
\end{figure}

\begin{table*}[t]
\small
\renewcommand\arraystretch{1.1}
\caption{\reviseref{} Testing results of five perception systems with with different guidance approaches.}
\label{tab:rq_fault_detection}
\begin{tabular}{lllccccclccccc}
\toprule
                                                         & \multicolumn{1}{c}{}                                    &  & \multicolumn{5}{c}{\textbf{3D Object Detection}} &  & \multicolumn{5}{c}{\textbf{2D Object Detection}} \\ \cmidrule{4-8} \cmidrule{10-14} 
\multirow{-2.4}{*}{\textbf{Metric}}                        & \multicolumn{1}{c}{\multirow{-2.4}{*}{\textbf{Method}}} &  & EPNet  & FConv  & CLOCs & $\star$ Second & Avg.  &  & EPNet  & FConv & CLOCs & $\star$ CasRCNN & Avg.  \\
\midrule
                                                         & \reviseref{}MetaOD                                                  &  & \reviseref{}0.05   & \reviseref{}1.84   & \reviseref{}0.14  & \reviseref{}---            & \reviseref{}0.68  &  & \reviseref{}0.01   & \reviseref{}1.21  & \reviseref{}0.12  & \reviseref{}1.12            & \reviseref{}0.62  \\
                                                         & \reviseref{}TauLim                                                  &  & \reviseref{}0.14   & \reviseref{}2.40   & \reviseref{}0.28  & \reviseref{}5.78           & \reviseref{}2.15  &  & \reviseref{}0.01   & \reviseref{}0.63  & \reviseref{}0.16  & \reviseref{}---             & \reviseref{}0.27  \\
                                                         & \reviseref{}MetaOD+TauLim                                           &  & \reviseref{}0.17   & \reviseref{}4.36   & \reviseref{}2.02  & \reviseref{}5.86           & \reviseref{}3.10  &  & \reviseref{}0.01   & \reviseref{}1.49  & \reviseref{}1.28  & \reviseref{}1.05            & \reviseref{}0.96  \\
                                                         & {\tool} (Random)                                                  &  & 0.52   & 20.34  & 5.77  & 9.15           & 8.95  &  & 1.91   & 16.89 & 6.17  & 20.98           & 11.49 \\
\multirow{-5}{*}{AP Difference $\uparrow$}               & {\tool} (Guided)                                                   &  & 10.25  & 32.31  & 22.64 & 26.01          & 22.80 &  & 2.84   & 24.29 & 16.64 & 30.09           & 18.47 \\
\rowcolor[HTML]{EFEFEF} 
\cellcolor[HTML]{EFEFEF}                                 & \reviseref{}MetaOD                                                  &  & \reviseref{}1      & \reviseref{}17     & \reviseref{}7     & \reviseref{}---            & \reviseref{}8     &  & \reviseref{}2      & \reviseref{}32    & \reviseref{}18    & \reviseref{}35              & \reviseref{}22    \\
\rowcolor[HTML]{EFEFEF} 
\cellcolor[HTML]{EFEFEF}                                 & \reviseref{}TauLim                                                  &  & \reviseref{}2      & \reviseref{}15     & \reviseref{}7     & \reviseref{}16             & \reviseref{}10    &  & \reviseref{}18     & \reviseref{}22    & \reviseref{}18    & \reviseref{}---             & \reviseref{}19    \\
\rowcolor[HTML]{EFEFEF} 
\cellcolor[HTML]{EFEFEF}                                 & \reviseref{}MetaOD+TauLim                                           &  & \reviseref{}2      & \reviseref{}19     & \reviseref{}8     & \reviseref{}17             & \reviseref{}12    &  & \reviseref{}19     & \reviseref{}21    & \reviseref{}17    & \reviseref{}25              & \reviseref{}21    \\
\rowcolor[HTML]{EFEFEF} 
\cellcolor[HTML]{EFEFEF}                                 & {\tool} (Random)                                                  &  & 9      & 83     & 25    & 23             & 35    &  & 8      & 50    & 21    & 56              & 34    \\
\rowcolor[HTML]{EFEFEF} 
\multirow{-5}{*}{\cellcolor[HTML]{EFEFEF}Location Error} & {\tool} (Guided)                                                  &  & 11     & 86     & 24    & 21             & 36    &  & 7      & 70    & 22    & 80              & 45    \\
                                                         & \reviseref{}MetaOD                                                  &  & \reviseref{}1      & \reviseref{}89     & \reviseref{}52    & \reviseref{}---            & \reviseref{}47    &  & \reviseref{}0      & \reviseref{}2     & \reviseref{}1     & \reviseref{}2               & \reviseref{}1     \\
                                                         & \reviseref{}TauLim                                                  &  & \reviseref{}15     & \reviseref{}60     & \reviseref{}33    & \reviseref{}56             & \reviseref{}41    &  & \reviseref{}3      & \reviseref{}2     & \reviseref{}1     & \reviseref{}---             & \reviseref{}2     \\
                                                         & \reviseref{}MetaOD+TauLim                                           &  & \reviseref{}14     & \reviseref{}62     & \reviseref{}37    & \reviseref{}61             & \reviseref{}43    &  & \reviseref{}2      & \reviseref{}3     & \reviseref{}1     & \reviseref{}2               & \reviseref{}2     \\
                                                         & {\tool} (Random)                                                  &  & 10     & 65     & 38    & 67             & 45    &  & 0      & 1     & 1     & 1               & 1     \\
\multirow{-5}{*}{False Detection}                        & {\tool} (Guided)                                                  &  & 9      & 81     & 39    & 71             & 50    &  & 0      & 4     & 1     & 3               & 2     \\
\rowcolor[HTML]{EFEFEF} 
\cellcolor[HTML]{EFEFEF}                                 & \reviseref{}MetaOD                                                  &  & \reviseref{}0      & \reviseref{}12     & \reviseref{}7     & \reviseref{}---            & \reviseref{}7     &  & \reviseref{}0      & \reviseref{}4     & \reviseref{}3     & \reviseref{}4               & \reviseref{}3     \\
\rowcolor[HTML]{EFEFEF} 
\cellcolor[HTML]{EFEFEF}                                 & \reviseref{}TauLim                                                  &  & \reviseref{}4      & \reviseref{}18     & \reviseref{}7     & \reviseref{}1              & \reviseref{}8     &  & \reviseref{}0      & \reviseref{}2     & \reviseref{}2     & \reviseref{}---             & \reviseref{}2     \\
\rowcolor[HTML]{EFEFEF} 
\cellcolor[HTML]{EFEFEF}                                 & \reviseref{}MetaOD+TauLim                                           &  & \reviseref{}4      & \reviseref{}23     & \reviseref{}10    & \reviseref{}1              & \reviseref{}10    &  & \reviseref{}0      & \reviseref{}2     & \reviseref{}2     & \reviseref{}2               & \reviseref{}1     \\
\rowcolor[HTML]{EFEFEF} 
\cellcolor[HTML]{EFEFEF}                                 & {\tool} (Random)                                                  &  & 31     & 135    & 90    & 34             & 72    &  & 5      & 75    & 69    & 75              & 56    \\
\rowcolor[HTML]{EFEFEF} 
\multirow{-5}{*}{\cellcolor[HTML]{EFEFEF}Object Missing} & {\tool} (Guided)                                                  &  & 49     & 178    & 153   & 51             & 108   &  & 12     & 96    & 121   & 90              & 80   
\\\bottomrule
\end{tabular}
\end{table*}

Fig.~\ref{fig: human_val} shows the participants' preference among four different data synthesis pipelines in terms of the \textit{image naturalness}, \textit{point cloud naturalness}, and \textit{modality consistency} between image and point cloud. We find that more participants prefer {\tool} over the three baseline methods on all three degrees. Wilcoxon's rank-sum test results suggest that the mean ranking differences between {\tool} and the second-chosen method are statistically significant ($p$<0.0001). 
These results suggest that our autonomous driving practitioners agree that {\tool} provides the most realistic and natural synthesized data among all four pipelines.

\subsection{RQ2: Fault Detection Capability}

Table~\ref{tab:rq_fault_detection} shows the testing results of five perception systems on 2D/3D object detection tasks. A larger difference in average precision (AP) signifies that a perception system performs considerably worse on the generated test cases compared to the original dataset. 
\reviseline{Compare with other baselines}, we observe a \reviseline{clear and} consistent trend of decreased AP performance in {\tool} across various tasks and perception systems. 
These results affirm that {\tool} is effective in generating critical and challenging test cases.
Furthermore, our proposed method proves to be proficient in detecting various categories of errors, particularly in identifying \textit{object missing} errors. However, it is worth noting an exception, where only a small number of false detection errors were detected for 2D object detection. This might be due to the relatively small image size in KITTI compared with the point cloud. As a result, cases where a detected bounding box is disjoint from all ground truth bounding boxes (i.e., IOU=0) are rare.


We also find that compared with the random baseline, our fitness-guided testing leads to a more significant decrease in average precision (AP) compared to the random testing baseline. This finding confirms the effectiveness of {\tool}'s fitness metric and its guidance, establishing that {\tool} is more efficient at generating test cases. Additionally, we observe that {\tool} triggered a higher number of object missing errors compared to the random testing baseline. These results suggest that {\tool}'s fitness-guided strategy improves overall testing efficiency, resulting in the synthesis of more error-revealing test cases.
 


\subsection{RQ3: Performance Improvement}

\begin{table}[t]
\renewcommand\arraystretch{1.1}
\setlength{\tabcolsep}{1.1pt}
\footnotesize
\caption{\reviseref{} Five perception system's performance after retraining with different approaches on the generated tests.}
\label{tab:rq_retrain}
\begin{tabular}{llcccccc}
\toprule
\textbf{Model}                                  & \textbf{Det.} & \multicolumn{1}{l}{\textbf{\makecell[c]{\tool  \\ (Guided)}}} & \multicolumn{1}{l}{\textbf{\makecell[c]{\tool  \\ (Random)}}} & \multicolumn{1}{l}{\reviseref{}\textbf{MetaOD}} & \reviseref{}\textbf{TauLim} & \multicolumn{1}{l}{\textbf{\makecell[c]{\reviseref{}MetaOD \\ \reviseref{}+TauLim}}} & \multicolumn{1}{l}{\textbf{Original}} \\
\midrule
\multirow{2}{*}{EPNet} & 3D                 & \reviseref{}80.87  & \reviseref{}77.34  & \reviseref{}75.51  & \reviseref{}74.37 & \reviseref{}75.39          & \reviseref{}75.53    \\
                       & 2D                 & \reviseref{}89.73  & \reviseref{}88.88  & \reviseref{}89.04  & \reviseref{}88.77 & \reviseref{}88.80          & \reviseref{}89.11    \\
\rowcolor[HTML]{EFEFEF} \cellcolor[HTML]{EFEFEF}
 & 3D                 & \reviseref{}82.72  & \reviseref{}79.30  & \reviseref{}49.34  & \reviseref{}65.19 & \reviseref{}49.88          & \reviseref{}57.98    \\
\rowcolor[HTML]{EFEFEF} \multirow{-2}{*}{\cellcolor[HTML]{EFEFEF}FConv}                       & 2D                 & \reviseref{}90.63  & \reviseref{}88.30  & \reviseref{}62.46  & \reviseref{}71.90 & \reviseref{}63.88          & \reviseref{}70.56    \\
\multirow{2}{*}{CLOCs} & 3D                 & \reviseref{}83.63  & \reviseref{}81.47  & \reviseref{}68.73  & \reviseref{}69.79 & \reviseref{}70.74          & \reviseref{}62.34    \\
                       & 2D                 & \reviseref{}90.49  & \reviseref{}89.19  & \reviseref{}80.36  & \reviseref{}80.31 & \reviseref{}80.29          & \reviseref{}71.10    \\
\rowcolor[HTML]{EFEFEF}
$\star$ Second         & 3D                 & \reviseref{}76.43  & \reviseref{}74.89  & \reviseref{}---                                 & \reviseref{}71.04 & \reviseref{}71.51          & \reviseref{}65.78    \\
$\star$ CasRcnn        & 2D                 & \reviseref{}94.86  & \reviseref{}93.20  & \reviseref{}60.03  & \reviseref{}---                                & \reviseref{}59.82          & \reviseref{}60.21    \\
\midrule
\multicolumn{2}{c}{Average AP}              & \reviseref{}86.17  & \reviseref{}84.07  & \reviseref{}69.35  & \reviseref{}74.48 & \reviseref{}70.04          & \reviseref{}69.08   \\

\bottomrule
\end{tabular}
\end{table}

Table~\ref{tab:rq_retrain} shows the AP performance of all subject perception systems after retraining with test cases generated by {\tool} or \reviseline{other baseline methods.
{\tool} can significantly improve the performance of all systems by retraining. }
This result indicates that the data generated by {\tool} includes challenging cases that can significantly help improve a perception system's robustness, showing the effectiveness of our testing. 
\reviseline{Compared with {\tool}, other baseline methods show relatively low capability in improving performance and in some cases can even be harmful. 
A possible reason is that test cases generated by single-sensor testing methods or their combination fail to maintain modal consistency and therefore hardly benefit in the retraining process.}

Moreover, we find that the average AP performance improvement with the fitness-guided testing and random testing across different systems are 86.05 and 83.79, respectively. Though both testing strategies can synthesize data that potentially help improve a perception system's performance, our fitness-guided testing is specifically more effective and efficient. This might largely attribute to the fact that {\tool}'s fitness-guided testing can generate more error-revealing test cases.

\section{Discussion}


{\bf Data realism affects testing effectiveness.}
{\tool} is designed to automatically generate realistic multi-modal data to assess the potential risks of a perception system when deployed in real-world scenarios. Both our quantitative evaluation and user study results confirm the realism of the data synthesized by {\tool}. Experiments with five perception systems further suggest that realistic data facilitates effective and efficient testing. We attribute {\tool}'s superior performance to two important designs.

Firstly, our physical-aware object insertion can generate natural yet challenging multi-modal test data for a perception system. Existing mutation-based testing methods usually face challenges that the perturbation (e.g., adversarial noises) might not be naturally exist in the real world. Therefore, it is questionable if the error-revealing test cases generated by these methods could help understand a SUT's potential risks. Compared with existing object insertion testing methods, {\tool} takes real-world physical constraints into account to ensure the semantic validity of its insertion. 

\sloppy Secondly, {\tool} simultaneously generates modality-consistent data across different sensors, which is crucial for testing MSF-based perception systems. Existing testing methods usually focus on single-sensor perception systems. Indeed, a combination of different single-sensor testing methods can also be used for testing an MSF system. However, without ensuring modality-consistency, the validity of the generated test cases becomes questionable. For instance, mis-aligned pairs of images and point clouds are rarely found in real-world autonomous driving applications. 

\vspace{1mm}
{\noindent \bf Testing efficiency trade-offs.}
Different parameter configurations could potentially affect {\tool}'s testing efficiency. Intuitively, inserting more objects could create more challenging test cases and potentially reveal more system faults. However, as the number of insertions grows, it might become harder for {\tool} to find possible positions for the insertion. We set up an ablation study to verify {\tool}'s testing efficiency trade-offs.
Specifically, We set the maximum number of trials $TRY\_NUM$ as 10 and the maximum object insertion number $N$ from 1 to 6. 
We randomly select 50 test seeds from KITTI and record the proportion of seeds that successfully generate test cases with different $N$ and calculate the average number of iterations required per seed.

\begin{table}[t]
\caption{Number of successful seeds and average of iterations for insertion in different numbers of inserted objects.}
\label{tab: ablation}
\renewcommand\arraystretch{1.1}
\setlength{\tabcolsep}{6pt}
\footnotesize
\begin{tabular}{llrrrrrr}
\toprule
\multirow{2}{*}{\textbf{Model}}   & \multicolumn{1}{l}{\multirow{2}{*}{\textbf{Metric}}} & \multicolumn{6}{c}{\textbf{Number of inserted objects: $N$}} \\
\cmidrule{3-8}
                         & \multicolumn{1}{c}{}                        & \multicolumn{1}{c}{1}     & \multicolumn{1}{c}{2}     & \multicolumn{1}{c}{3}      & \multicolumn{1}{c}{4}     & \multicolumn{1}{c}{5}     & \multicolumn{1}{c}{6}     \\
\midrule
\multirow{2}{*}{EPNet}   & Seeds      & 0.70 & 0.48 & 0.28  & 0.20  & 0.14  & 0.12  \\
                         & Iterations & 1.46 & 6.42 & 16.93 & 27.70 & 43.57 & 53.00 \\
                         \cellcolor{lightgray}& \cellcolor{lightgray}Seeds   & \cellcolor{lightgray}0.92 & \cellcolor{lightgray}0.76 & \cellcolor{lightgray}0.60  & \cellcolor{lightgray}0.50  & \cellcolor{lightgray}0.40  & \cellcolor{lightgray}0.30  \\
\cellcolor{lightgray}\multirow{-2}{*}{FConv}   & \cellcolor{lightgray}Iterations  & \cellcolor{lightgray}1.00 & \cellcolor{lightgray}3.58 & \cellcolor{lightgray}7.30  & \cellcolor{lightgray}11.20 & \cellcolor{lightgray}16.90 & \cellcolor{lightgray}25.87\\
\multirow{2}{*}{CLOCs}   & Seeds                                       & 0.74  & 0.66  & 0.48   & 0.34  & 0.24  & 0.22  \\
                         & Iterations                                  & 2.03  & 4.94  & 11.25  & 20.53 & 34.92 & 41.45 \\
                         \cellcolor{lightgray}& \cellcolor{lightgray}Seeds                                       & \cellcolor{lightgray}0.84  & \cellcolor{lightgray}0.72  & \cellcolor{lightgray}0.54   & \cellcolor{lightgray}0.40   & \cellcolor{lightgray}0.32  & \cellcolor{lightgray}0.20   \\
\cellcolor{lightgray}\multirow{-2}{*}{$\star$ Second} & \cellcolor{lightgray}Iterations                                  & \cellcolor{lightgray}1.29  & \cellcolor{lightgray}4.56  & \cellcolor{lightgray}10.89  & \cellcolor{lightgray}19.65 & \cellcolor{lightgray}28.25 & \cellcolor{lightgray}50.20 \\
\multirow{2}{*}{$\star$ CasRCNN} & Seeds                                       & 0.88  & 0.74  & 0.56   & 0.48  & 0.32  & 0.26  \\
                         & Iterations                                  & 1.45  & 4.73  & 10.25  & 14.71 & 27.00 & 36.54 \\
\bottomrule
\end{tabular}
\end{table}

Table~\ref{tab: ablation} validates our hypothesis. As the number of objects increases, the proportion of successful seeds decreases. About 50\% of the seeds are able to insert three objects, and this value drops to about 25\% when N is 6.
Moreover, the average number of iterations grows exponentially with the number of inserted objects. We further qualitatively check the seeds and find that the complex scenes in autonomous driving (e.g., cars driving on a busy city road) could significantly increase the difficulty of object insertion. 
In this paper, we set $N$ is 3 to balance the testing efficiency trade-offs.
\vspace{1mm}
{\noindent \bf Limitations and future work.} \reviseline{Our {\tool} is able to generate test cases to reveal three different types of faults in object detection. In future work, one may consider controlling the coefficients in Eq.~\ref{eq: fitness} to generate test cases with preferences for different fault types according to the developer's intention.} Currently, {\tool} leverages a model, S$^{2}$CRNet~\cite{liang2022spatial}, to improve the naturalness of the synthesized images. It would be worth-while to explore more advanced methods, e.g., SOTA generative AI and diffusion models~\cite{ramesh2022hierarchical, rombach2022high}. Furthermore, we have only experimented {\tool} with object detection systems. One can consider how to leverage {\tool} to test different MSF perception systems, e.g., object tracking systems. Compared with testing object detection, testing object tracking would require the synthesis of sequential data (i.e., a series of temporally correlated data). How to ensure the temporal consistency between data frames and how to generate trajectories for the object to be inserted could be challenging but worth investigating.

\vspace{1mm}
{\noindent \bf Threats to validity.}
In terms of \textit{construct validity}, one potential threat comes from the measurement of the realism of {\tool}'s generated data since there is no ground truth label. To mitigate this, we conduct both quantitative and qualitative experiments to assess the quality of the synthesized data. Another construct threat lies in the randomness inherent in our experiments, specifically in the testing (RQ2) and retraining (RQ3). To combat this, we repeat our experiment on testing and retraining for five times to reduce the influence of randomness.


In terms of \textit{internal validity}, one potential threat is that the data generated by the virtual simulator may differ from real-world. Besides, the quality of 3D models in {\tool}'s object database could also affect the quality of synthesized data. To mitigate these, we utilize the well-known simulators (i.e., Blender and Open3D) and select a popular 3D object database ShapeNet.

In terms of \textit{external validity}, one potential threat is that our analysis results may not be generalized to other perception systems. To combat this threat, we experimented with three MSF systems with different fusion mechanisms and two single-sensor detection models to evaluate the effectiveness of {\tool}.

\section{Related Works}

\noindent\textbf{Multi-sensor Fusion.}
MV3D~\cite{chen2017multi} is a pioneering work in the field of AI-enabled multi-sensor fusion. 
It introduces a multi-view based deep fusion framework that extracts hidden features from different view representations of 3D point clouds and images, enabling region-wise feature fusion.
To further enhance system performance, subsequent deep-fusion based methods directly fuse the features extracted from raw data to mitigate information loss caused by point cloud view transformations. For instance, MVX-net~\cite{sindagi2019mvx}, a voxel-based fusion method, and EPNet~\cite{huang2020epnet}, a point-based fusion method, are notable examples of such techniques.
In the realm of late fusion, CLOCs~\cite{pang2020clocs} is a representative work that fuses the output results of different detectors. 
DFMOT~\cite{wang2022deepfusionmot} is another late-fusion based work for object tracking that employs a four-level deep association mechanism to achieve a fast fusion of 2D and 3D trajectories.
Weak fusion often involves using 2D proposals as guidance to extract the frustum region from the point cloud~\cite{qi2018frustum,wang2019frustum}. F-PointNets~\cite{qi2018frustum} pioneered this approach for object detection, and FConv~\cite{wang2019frustum} extended it further by incorporating a post-grouping aggregation scheme, leading to end-to-end estimation and improved performance.

\vspace{1mm}
\noindent\textbf{Quality Assurance of Perception Systems.}
The first group of related works is on automatic testing for visual perception systems ~\cite{wang2020metamorphic,xie2022towards,zhang2021deepbackground}.
Wang et al. first propose MetaOD~\cite{wang2020metamorphic} to test object detection systems by inserting objects based on metamorphic testing theory.
However, this approach faces limitations when it comes to inserting image object instances in valid positions due to the lack of 3D information.
Another set of related work focuses on LiDAR-based perception systems robustness testing~\cite{guo2022lirtest,10172508,lin2022taulim}.
Guo et al.~\cite{guo2022lirtest} and Christian et al.~\cite{10172508} leverage data mutation operators to generate test point clouds and check the failure in perception systems based on metamorphic relationships. However, extending these testing methods to MSF perception systems becomes challenging due to the multi-modal data required by MSF systems.

On the other hand, there are a few works that focus on benchmarking MSF systems~\cite{zhong2022detecting,gao2023benchmarking}.
Recently, Gao et al.~\cite{gao2023benchmarking} create an early public benchmark of MSF systems and perform a large-scale empirical study to investigate their robustness performance.
Along this direction, we design and implement {\tool} for automated testing of MSF perception systems. We adopt the MSF system designed for object detection tasks provided by the MSF benchmark as our experimental subject.
There are also some other works that assess a perception system's quality from the security perspective by inserting adversarial objects~\cite{cao2021invisible,tu2021exploring,abdelfattah2021towards}. 

\vspace{1mm}
\noindent\textbf{General Deep Learning System Testing.}
Inspired by the effectiveness of code coverage applied to traditional software testing, 
researchers have proposed several neuron coverage criteria to guide the testing process of Deep Learning (DL) systems.
Pei et al.~\cite{pei2017deepxplore} first propose the neuron coverage criterion to measure the extent of state exploration in deep neural networks.
Ma et al.~\cite{ma2018deepgauge} further refined the neuron coverage criterion and proposed a set of fine-grained testing criteria.
With the guidance of the neuron coverage criterion, several test generation techniques have been proposed to detect erroneous behaviors of DNNs by increasing the neuron coverage~\cite{xie2019deephunter, du2019deepstellar}.
To detect erroneous behaviors of DL systems effectively, several domain-specific guidance metrics are proposed~\cite{feng2020deepgini,gao2022adaptive}.
DeepGini leverages Gini impurity to measure the fault-revealing capabilities of test cases~\cite{feng2020deepgini}.
In contrast, our work takes a different approach by focusing on a context-specified problem, specifically on the testing of real-world MSF perception systems. This specialized focus allows us to address the unique challenges and requirements posed by the systems under test.

\section{Conclusion}
In this paper, we present {\tool}, the first automated testing tool designed specifically for MSF perception systems. {\tool} employs a physical-aware approach to render modality-consistent object instances using virtual sensors. Then, {\tool} synthesizes realistic images and point clouds by inserting object instances into valid yet challenging positions within the target scene.
Moreover, {\tool} incorporates fitness metric guidance to boost the testing efficiency and effectiveness. We evaluate the performance of {\tool} using three state-of-the-art MSF-based detectors and two single-sensor based detectors.
The experimental results demonstrate that {\tool} efficiently detects erroneous behavior in the systems under test and improves a system's robustness through retraining.
In the future, we plan to extend the capabilities of {\tool} to support a wider range of MSF perception systems and tasks, such as object tracking and depth completion. By doing so, we aim to broaden its applicability and impact in the field of multi-sensor fusion system testing.

\section*{Acknowledgement}
We would like to thank anonymous reviewers for their constructive comments and feedback. This project was partially funded by the National Natural Science Foundation of China under Grant No.61932012, No.61832009 and No.62002158. 
This work was also supported in part by Canada CIFAR AI Chairs Program, the Natural Sciences and Engineering Research Council of Canada (NSERC No.RGPIN-2021-02549, No.RGPAS-2021-00034, No.DGECR-2021-00019); JST-Mirai Program Grant No.JPMJMI20B8, JSPS KAKENHI Grant No.JP21H04877, No.JP23H03372; as well as the support from TIER IV, Inc. and Autoware Foundation.

\bibliographystyle{ACM-Reference-Format}
\bibliography{reference}

\end{document}